# Energy Storage Autonomy in Renewable Energy Systems Through Hydrogen Salt Caverns

D. Franzmann[1,2,*], T. Schubert[1], H. Heinrichs[1], P. A. Kukla[3], D. Stolten[1,2]

1 Forschungszentrum Jülich GmbH, Institute of Energy and Climate Research – Jülich Systems Analysis (ICE-2), 52425 Jülich, Germany

2 RWTH Aachen University, Chair for Fuel Cells, Faculty of Mechanical Engineering, 52062 Aachen, Germany

3 RWTH Aachen University, Chair of Geology and Sedimentary Systems and Geological Institute, Faculty of Georesources and Materials Engineering, 52062 Aachen, Germany

* corresponding author: d.franzmann@fz-juelich.de

## Abstract

The expansion of renewable energy sources leads to volatility in electricity generation within energy systems. Subsurface storage of hydrogen in salt caverns can play an important role in long-term energy storage, but their global potential is not fully understood. This study investigates the global status quo and how much hydrogen salt caverns can contribute to stabilizing future renewable energy systems. A global geological suitability and land eligibility analysis for salt cavern placement is conducted and compared with the derived long-term storage needs of renewable energy systems. Results show that hydrogen salt caverns can balance between 43% and 66% of the global electricity demand and exist in North America, Europe, China, and Australia. By sharing the salt cavern potential with neighboring countries, up to 85% of the global electricity demand can be stabilized by salt caverns. Therefore, global hydrogen can play a significant role in stabilizing renewable energy systems.

A massive expansion of renewable energy is needed to combat climate change [1]. However, the variability of renewable energy sources such as wind and PV requires long-term storage over months to years [2, 3]. Today, the volatility of energy systems is backed up by fossil plants [4], which will not be available in 100% decarbonized energy systems. Therefore, alternative storage options such as renewable hydrogen are needed [5]. The required storage capacity of a renewable energy system in 2050 is expected to be as high as 1290 TWh worldwide [6]. Large scale energy storage is also needed to increase the national energy security, as seen in Europe's dependency on 1,135 TWh of natural gas storages after the gas import stop during the Russo-Ukrainian war in 2022 [7].

Underground hydrogen storage is a promising option for large storage need [8, 9]. It offers an economic advantage over compressed air energy storage [10], provides a huge storage potential compared to pumped hydro storage, which is geographically limited and often already depleted [10], and does not compete for land use like biomass-based storage. Hydrogen can be stored underground in salt caverns or porous rocks such as depleted gas fields or deep saline aquifers [8, 11]. Among all the options, hydrogen salt caverns are preferred due to several advantages: (1) they are technically mature as they are stable for decade s[12, 13], (2) provide a high tightness and self-sealing properties for hydrogen usage [12, 14], (3) have low interactions with hydrogen [14], (4) require less cushion gas than other underground storages [15] and (5) have a high cyclic flexibility and large storage volume which makes them ideal for their use in volatile renewable energy systems [12, 16]. The leaching and storage of natural gas in salt caverns has been a well-established technology since the 1960s [17] with more than 84 caverns in operation [18]. Between 1972 and 2014, four commercial salt caverns were commissioned for hydrogen use [18], and several projects for new hydrogen caverns are currently funded [19]. In general, hydrogen salt caverns are technically mature with only few risks such as geochemical interactions, leaching issues, and hydrogen diffusion that are already known in the industry [20]. In conclusion, hydrogen salt cavern storage is currently the most favored solution for seasonal storage in renewable energy systems [21].

However, the magnitude and spatial distribution of the technical storage potential of hydrogen salt caverns are not fully understood. This understanding is crucial for determining where energy system designers can rely on seasonal storage from hydrogen salt caverns and where alternative solutions for seasonal balancing are needed to ensure the reliability of renewable energy systems. Suitability for cavern construction is determined based on geological criteria such as thickness and depth of the salt deposit, absence of interbeds, and complexity of the salt structure [15]. In addition to geological criteria, land surface criteria such as a minimum distance from settlements due to land subsidence must be considered [22]. The regionally largest study considering these criteria was conducted by Caglayan et al. [22], who focused on Europe and found a storage potential of 85 PWh. Other smaller studies investigate the local hydrogen salt cavern potentials of a German federal state with a capacity of 386 TWh [23], the UK with 2.15 PWh of storage [24], the Netherlands with 277 TWh [25], and selected salt domes and bedded salt in Poland [26, 27]. Qualitative studies without calculation of technical potentials also exist for parts of Canada [28, 29] and parts of Portugal [30]. Derakhshani et al. [31] propose a novel approach to estimate the feasibility of salt cavern storage, but only for a specific deposit in Poland. However, there is no global analysis of the storage potential of hydrogen salt caverns, as specifically pointed out by Hunt et al. [32].

The crucial need for seasonal hydrogen storage in hydrogen salt caverns within future renewable energy systems and the lack of knowledge about their global technical potential raises the following question: Where do potentials for hydrogen salt caverns exist and to what extent can they contribute to the seasonal balancing of renewable energy systems? We follow a three-step approach: First, the global hydrogen salt cavern potential is derived based on a geological analysis of salt deposits and by conducting a location specific land eligibility analysis for cavern placement worldwide. Second, the seasonal hydrogen storage demand for balancing renewable energy systems is derived globally on a national level. Finally, the new variable „sufficiency of hydrogen storage" is introduced to understand the extent to which salt caverns can contribute to balancing renewable energy systems in 2050.

# Main

## 1.1 Global hydrogen salt cavern potentials

The global technical capacity potential of hydrogen salt cavern storage is based on the geological analysis of 174 currently known groups of existing salt deposits around the world recognized by the Solution Mining Research Institute [18] and several national studies described in the Methodology section. Of the existing salt deposits, 31 salt deposits were identified as guaranteed suitable for cavern placement based on a geological suitability analysis for cavern construction. As salt deposits are not uniform in shape and not fully explored in detail, 34 salt deposits are found to be at least partially useable. The partially suitable deposits have areas that are suitable for cavern construction and areas that are not suitable for cavern construction, with the numeric value of the usable portion being geologically unexplored. This uncertainty is treated with two geological cases: the "guaranteed suitable" case as a lower bound of the geologically suitable and the "guaranteed and partly suitable" case as an upper bound (see Figure 1).

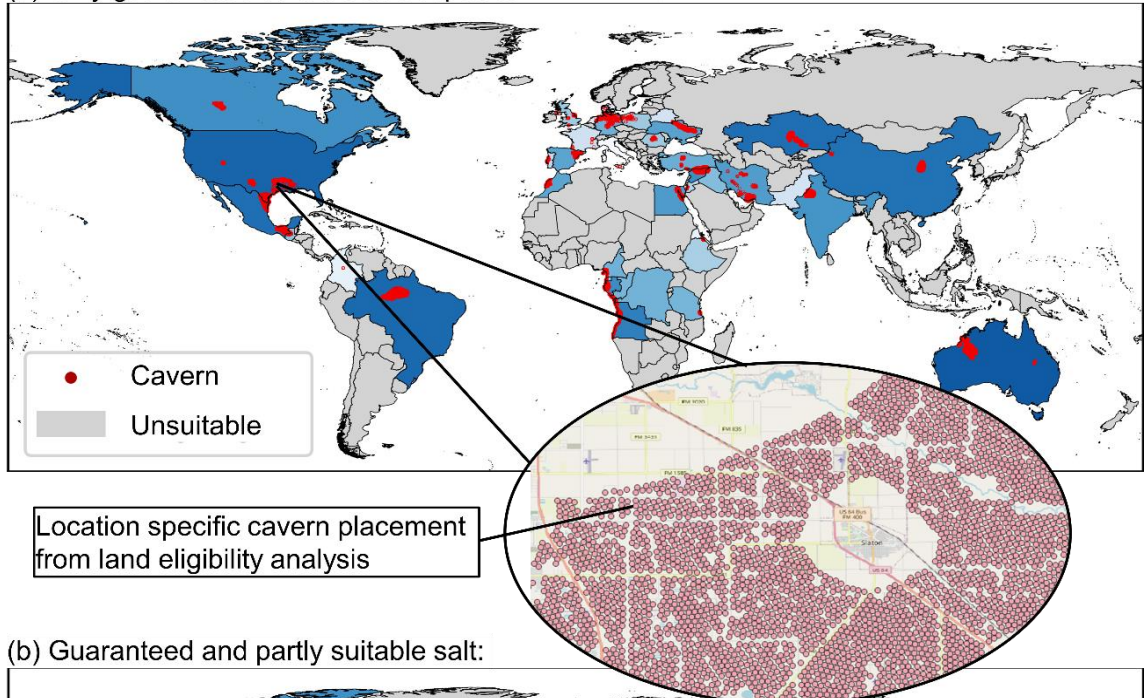

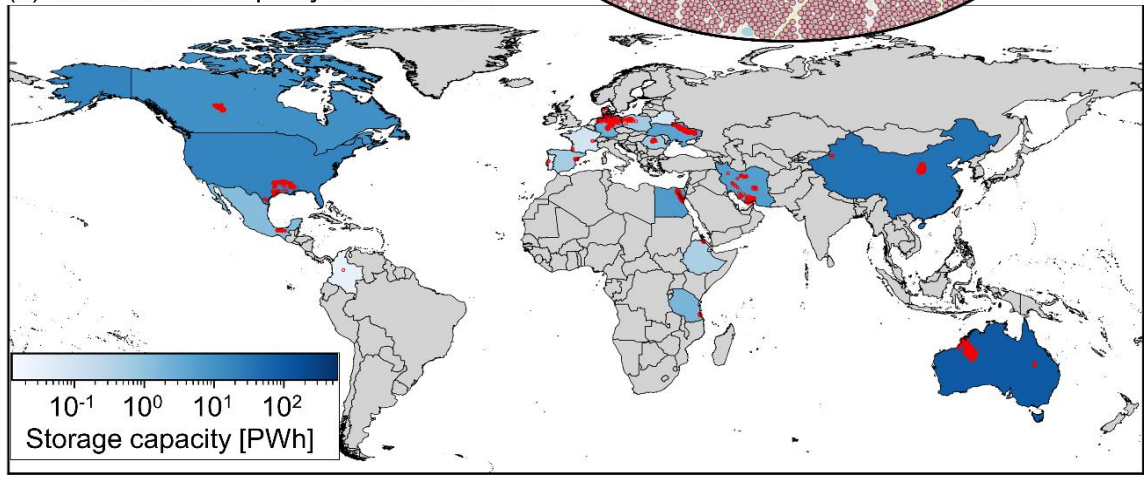

**Figure 1: Global potential for underground hydrogen salt cavern storage based on a location specific land eligibility analysis. Figure (a) shows the capacity of salt cavern hydrogen storage for guaranteed suitable**

**salt deposits (lower bound of capacity). Figure (b) shows the added capacity for guaranteed suitable and partially suitable salt deposits (upper bound of capacity). Background map © OpenStreetMap-Contributors**

Of the 12.84 million km² of horizontal salt area worldwide, 0.97 million km² of salt area is geologically suitable for hydrogen cavern construction and 2.50 million km² is partially suitable for hydrogen cavern construction. In total, 1.8% of the world's surface area is suitable for cavern placement based on the currently known geological existence of salt structures. Suitable and partially suitable salt structures can be found in every continent (see Figure 1).

However, not all geologically suitable surface areas are ultimately eligible for hydrogen cavern construction due to surface land restrictions. Approximately 54% of the surface area above salt structures cannot be used due to a minimum distance of 2,000 m from settlements for land subsidence, nature protection excludes 26% of the salt areas, and intact forests block 13.3% of the salt areas. Ultimately, only 0.80 million km² or 0.53% of the total land surface area is suitable for salt cavern construction (see red dots and areas in Figure 1). Globally, 197 PWh of the theoretical hydrogen storage potential is usable, while another 278 PWh is partially suitable. The largest guaranteed salt cavern potentials near global energy demand centers are in North America, Europe, China and Australia. In addition, the partially suitable hydrogen potential near future major demand centers can be found in India. Countries with a high energy demand and no potential for hydrogen salt cavern storage are Russia, Japan and Southeast-Asia. As shown in in Figure 1, the potential for hydrogen salt cavern storage is geographically concentrated but spread over different regions. In general, the technical storage potential in salt caverns exceeds the projected seasonal hydrogen storage demand within the IEA's net-zero scenario in 2050 [33] by two orders of magnitude. Therefore, self-sufficiency in seasonal energy storage is likely in regions with suitable salt cavern potential.

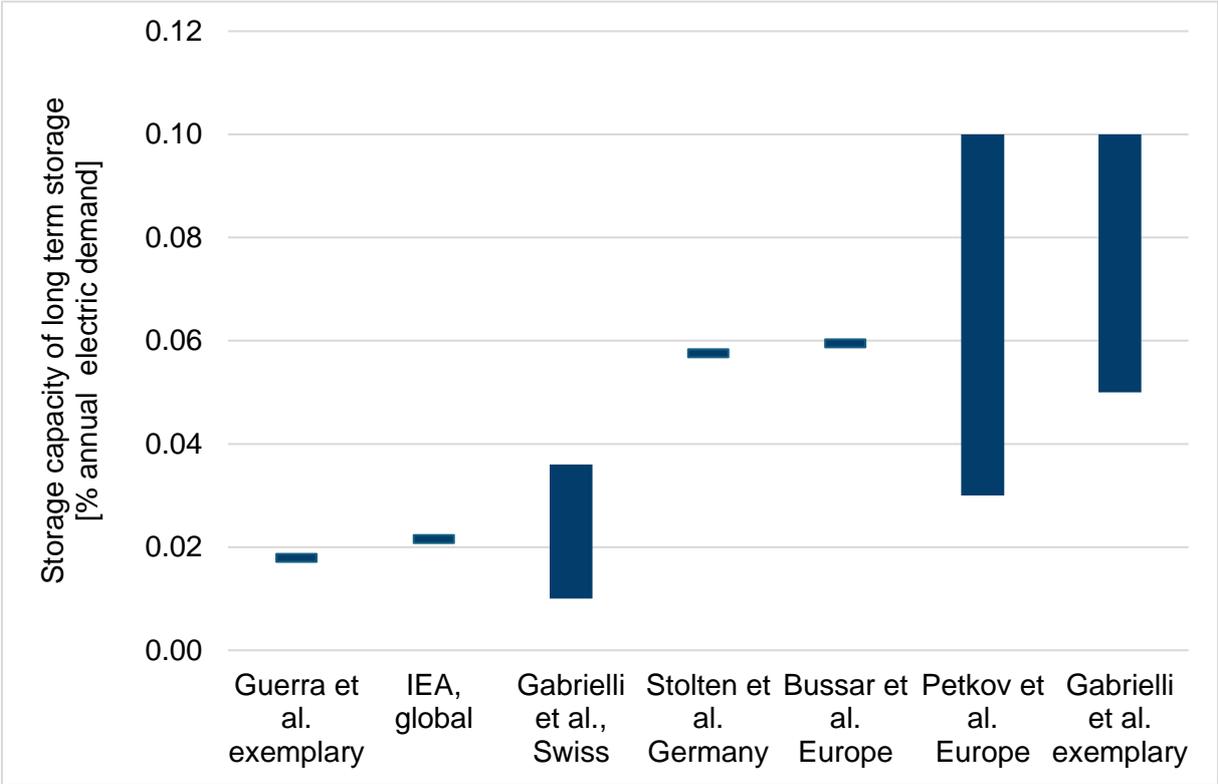

**Figure 2: Capacity of hydrogen long term storage in renewable energy scenarios for 2050 compared to the annual electricity demand [10, 34–38]**

**1.2 Hydrogen storage demand in renewable energy systems**

To calculate whether the hydrogen salt storage potential is sufficient as seasonal storage for renewable energy systems, the required size of seasonal storage in renewable energy systems must be estimated. Several studies have projected the required seasonal hydrogen storage for renewable energy systems in 2050 (see Figure 2). The studies indicate that the required size of salt cavern storage for stable, fully renewable energy systems is in a range of 2% to 10% of the annual electricity demand in 2050 of the respective studies (see Figure 2). Most of the variation is due to different assumptions even within the studies themselves. For instance, Gabrielli et al. [5] use different renewable capacity expansion scenarios in their studies, resulting in different sizes of hydrogen storages. Similarly, Petkov et al. [5] examine different climate zones for weather data, resulting in different hydrogen storages sizes ranging from 3% to 10% of the annual electricity demand. In addition, Stolten et al. [10] consider up to 9% of electricity generation in Germany in 2045 from dispatchable biomass and biomethane, reducing the required hydrogen storage to 6% of the annual electricity demand. In total, hydrogen storage of 10% of the annual electricity demand can be considered as the upper bound, guaranteeing stable renewable energy systems. Combined with the electricity demand from GCAM's global "Net-Zero 2050" scenario at the country level [39–41], the upper bound for the total required hydrogen storage size per country can be calculated as shown in the supplementary data.

**1.3 Sufficiency of hydrogen seasonal underground storage in salt caverns**

A comparison of the hydrogen storage capacities of the salt caverns (see Section 1.1) and the upper bound for the required hydrogen storage capacities (see Section 1.2) reveals the sufficiency of long-term storage based on hydrogen salt caverns (see Figure 3). The sufficiency of hydrogen storages is defined as:

$$Sufficiency\ of\ hydrogen\ storages := \frac{Storage\ potential_{region}}{Storage\ need_{region}}$$

It is a newly introduced metric that shows which countries can stabilize a future renewable energy system based on seasonal hydrogen storage on their own and which cannot.

The global hydrogen storage potential in salt caverns would exceed the seasonal storage needs of the whole world by a factor of more than 34 in the case of guaranteed geological suitability. Almost all countries with a potential for hydrogen storage in salt caverns have enough storage for their own needs, often exceeding their own demand by more than 1000%. With an average cavern lifetime of 60 years [13], hydrogen salt caverns can provide more than 600 years of seasonal hydrogen storage in these countries. In the guaranteed suitability case, up to 43% of the global electricity demand could be balanced. The highest sufficiency levels are observed in the USA (>2 262%), China (>3 817%), European countries (>1 123%), and Australia (161 563%), where hydrogen salt cavern storage shows significant potential. This suggests that in these regions, hydrogen salt caverns may be the most promising solution for balancing renewable energy fluctuations. Including partly suitable deposits, 66% of the global electricity demand could be balanced. Other large economies that are partial suitable for salt cavern construction are India (1 485%), Canada (7 105%), Brazil (32 727%) and the United Kingdom (673%). In these countries, there is a potential for hydrogen storage in salt cavern, but data currently available cannot describe the full extent of the potential. Therefore, further exploration activities are required to determine the suitable storage potential of salt cavern in these countries. Large economies without sufficient salt caverns are Japan, Russia, Indonesia and Italy with a sufficiency of only 17%. In these countries, other options for balancing the energy system need to be considered.

(a) Only guaranteed suitable salt deposits:

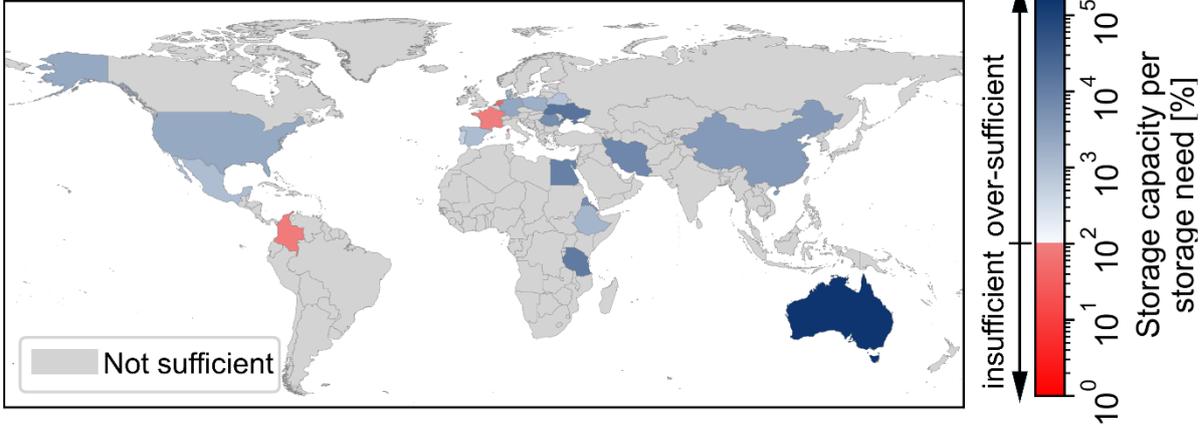

(b) Guaranteed and partly suitable salt deposits:

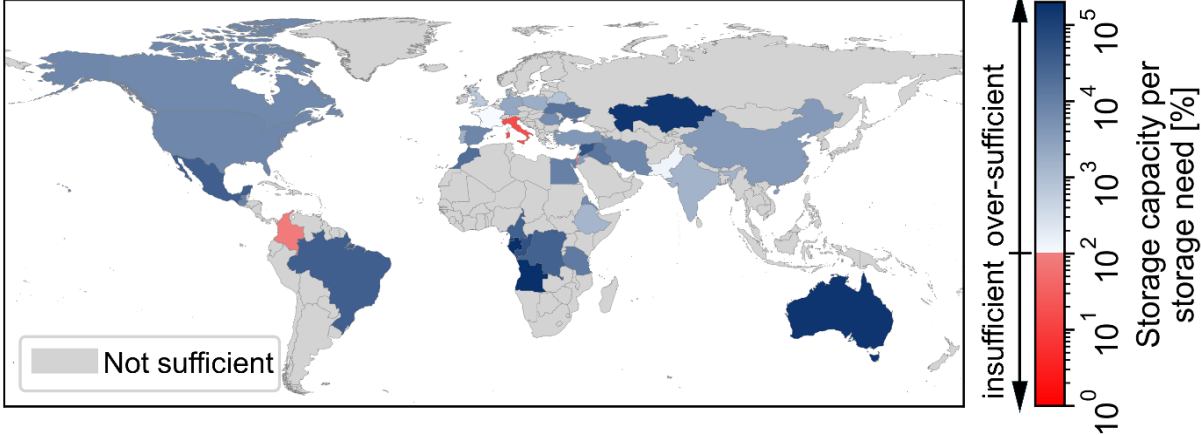

**Figure 3: Sufficiency of long-term energy storages from hydrogen salt caverns for renewable energy systems based on country wide usage. Figure (a) shows the guaranteed storage capacities and figure (b) the upper bound of hydrogen storage capacities. Blue countries are self-sufficient in terms of hydrogen storage, red and grey countries are not.**

## 1.4 Impact of cavern placement on sufficiency

Further technical options exist to increase the eligible storage potential: These include horizontal drilling, which enables the placement of caverns below protected areas, and variations in the placement spacing of caverns. Both options and the latter with a separation distance of either 3 or 5 times the diameter of a cavern, compared to the 4-diameter assumed in the base scenario, are investigated (see Figure 4). This is of particular interest for Colombia, France, Israel, Italy, Namibia and Palestine to enable self-sufficiency in hydrogen storage or for boosting the storage capacity globally. Among those options, decreasing the separation distance between caverns has the largest impact on a global scale (+ 72%). Horizontal drilling, which allows caverns to be below protected areas, changes the storage potential by 33% (see Figure 4). Consequently, Colombia and Israel would be able to meet their seasonal storage needs on their own, if either of the two options above were applied. However, this trend differs spatially.

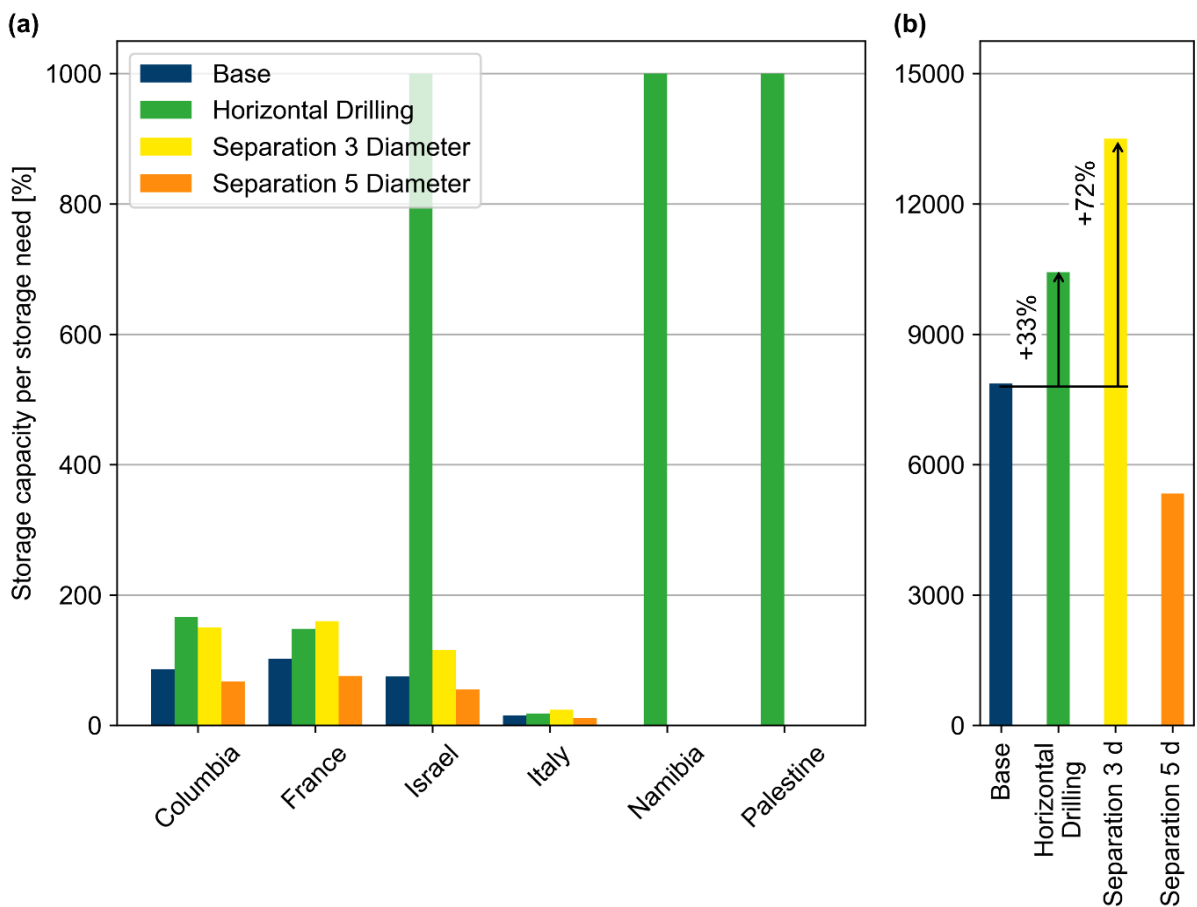

**Figure 4: Change of sufficiency of long-term hydrogen salt cavern storage for different salt cavern scenarios. For Columbia, Israel, Namibia and Palestine further technology research like horizontal drilling and lesser separation distance can enable sufficiency of hydrogen storages (Figure a). Globally, a smaller separation distance of salt caverns has the biggest impact on sufficiency (Figure b).**

Horizontal drilling enables Namibia and Palestine to utilize further hydrogen salt caverns potentials, increasing their sufficiency to 20 409% and 4 643%, respectively. Horizontal drilling has a high impact in these countries, as the salt potentials are located below the edge of nature protected areas. These areas can only be accessed for salt caverns placement using horizontal drilling technology. However, none of the investigated variants has a high absolute impact on the sufficiency in Italy. In France, with a lower separation distance of 3 diameters, the long-term storage hydrogen sufficiency can be increased by +60%. Therefore, horizontal drilling is a necessary technology for enhancing the storage capacity of each country in Figure 4. On a global scale, the separation distance has the greatest impact on salt cavern capacity.

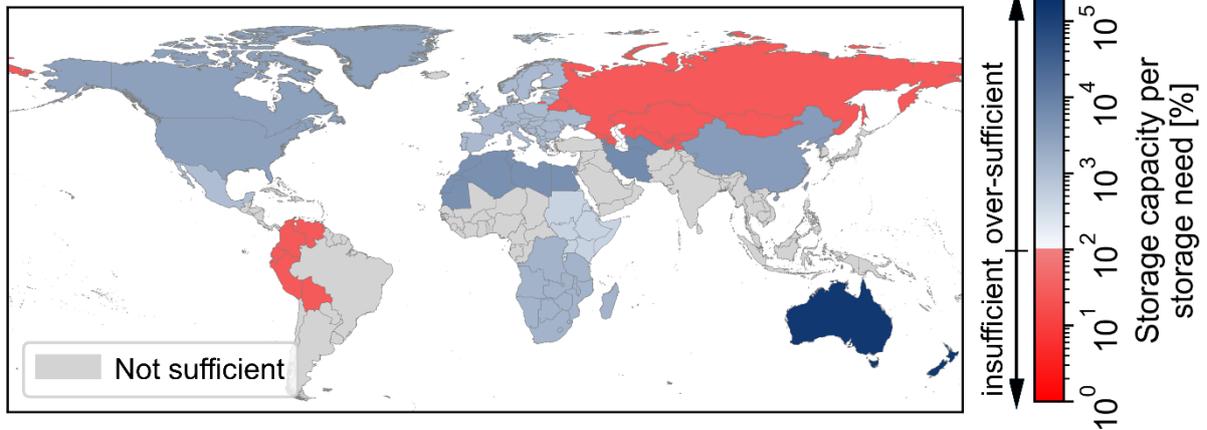

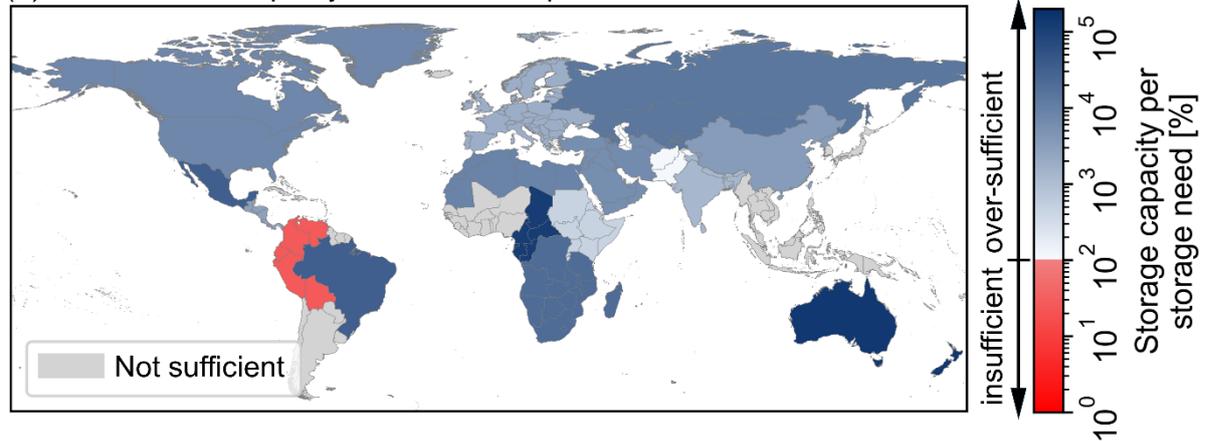

Figure 5: Sufficiency of long-term energy storages from hydrogen salt caverns if country are connected to regions. Figure (a) shows the guaranteed storage capacities and figure (b) the upper bound of hydrogen storage capacities. Blue countries are self-sufficient in terms of hydrogen storage, red and grey countries are not.

### 1.5 Impact of regional collaboration on sufficiency

An additional option for increasing sufficiency through hydrogen salt cavern storage is the shared usage of caverns through interconnected infrastructure with neighboring countries. Sharing hydrogen salt cavern capacities increases the number of countries with a sufficient amount of seasonal hydrogen storage from 17 to 102 out of 256 in the guaranteed suitability case and from 37 to 152 out of 256 in the partial suitability case. This contributes to an additional stabilization of more than 20% of the global electricity demand. In total, 59% in the guaranteed suitability case and 85% in the partial suitability case of the global energy demand can be balanced with hydrogen salt caverns. Interconnected regions that are guaranteed to have sufficient hydrogen salt cavern storage potentials by a factor of at least 10 are North America, Europe, North- and South Africa, China, Australia, and Iran. In addition, parts of Central Asia and Africa have sufficient hydrogen storage potential by a factor of 5-10. When partial suitable geological deposits are also considered, all regions except West Africa, Southeast Asia, Japan, western South America, and the Caribbean have sufficient seasonal hydrogen storage options. Therefore, currently known salt structures can help to balance up to 87% of the global electricity demand from renewable energy systems through seasonal hydrogen storage.

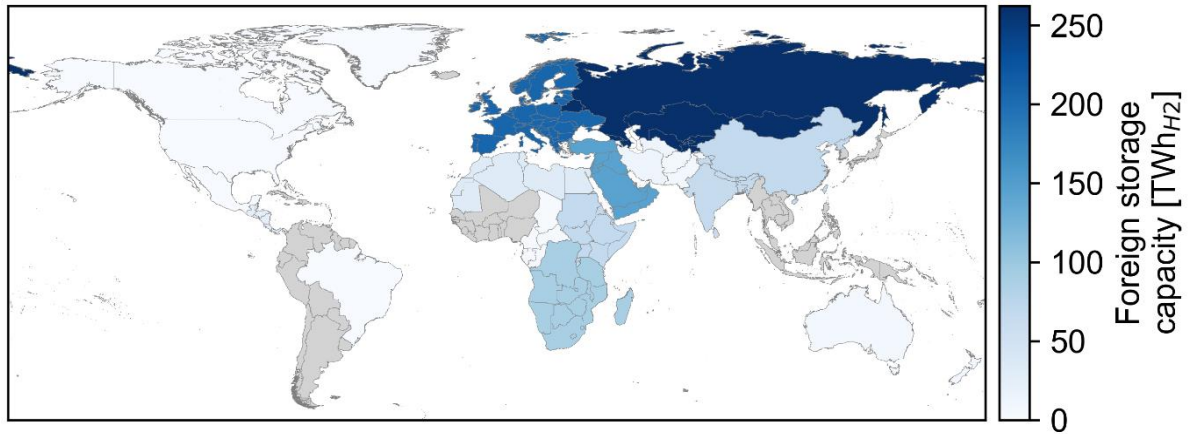

**Figure 6: Amount of storage capacity which is stored abroad within each global region to enable hydrogen storage of at least 10% electricity demand.**

However, the use of regionwide shared hydrogen storage potentials requires the interconnection of regions by hydrogen transport (see Figure 6). The hydrogen storage capacities used abroad could reach up to 262 TWh in North Asia due to the abundance of salt caverns in Russia. The storage abroad could be up to 207 TWh in Europe due to the regionally small country definition within Europe. This leads to an additional hydrogen transport demand. The operational time series of seasonal storages such as from Petkov et al. [5] show that hydrogen storage typically use one full cycle per year. Consequently, the energy content of the shared storage capacity must be transported at least once per year within each region. In contrast, the expected annual hydrogen trade within Europe in 2050 is about 1,325 TWh [42]. Therefore, the impact of storage on hydrogen transport is estimated to be +15% compared to European trade-based hydrogen transport.

## Discussion and Conclusions

A review of current the literature reveals that hydrogen salt cavern storage is the most promising option for seasonal storage. Going beyond existing literature, this paper demonstrates that hydrogen salt cavern storage is an available storage option in United States, Europe, China and Australia and most likely in Canada and India as well. For these regions, salt cavern storage represents a technologically mature and available technology that should be utilized for energy system stabilization. In total, up to 5,777 TWh of salt cavern storage may be needed for seasonal energy system balancing globally, from which up to 4,942 TWh can be built based on technical potentials. Until 2050, this equates to an average expansion of 198 TWh/a, which represents an annual increase of 202% of the current global gas cavern storage amount by working gas volume [43]. Although this represents a significant expansion for cavern construction industry, there are no substantial technical limitations to salt cavern construction, as it is a well-established technology for natural gas and hydrogen [18]. However, it is important to consider that cavern leaching results in the production of significant quantities of brine, which must be disposed of properly. Brine disposal into seawater as in Caglayan et al. [22] is not exhaustive, as this is often regulated due to its impact on marine ecosystems [44]. In contrast to disposal, further investigation of the potential synergies of brine production with the global salt industry may be beneficial. This study indicates that salt production from salt cavern leaching for hydrogen purposes could reach up to 1,191 Mt/a, which is a significant increase over the global salt production of 270 Mt/a in 2023 [45]. However, the additional disposal of brine needs to be further investigated beyond this study in order to exploit the technical storage potential of salt caverns worldwide.

In 85 to 155 out of 256 countries, primarily situated in Europe, Africa and Asia, the only possibility of getting access to salt cavern storage is through the utilization of neighboring countries potentials in combination with pipeline transportation of hydrogen. Technological solutions such as horizontal drilling and smaller separation of caverns have limited regional impact and are typically unable to increase the sufficiency of hydrogen storage. Since the only option for these 85 to 155 countries is the shared usage, international collaboration between countries is key to accessing cheap and large-scale hydrogen storage. The global increase in hydrogen transportation from using abroad salt cavern resources can be as high as 988 TWh/a globally. While long-distance transport of hydrogen is generally technically feasible and efficient [46], geopolitical factors as well as economic challenges of long-distance transportation of hydrogen may be a major barrier to international sharing of salt caverns.

No storage potential in salt caverns exists in Japan, Korea, Southeast Asia and East Africa, even if shared usage is considered. Additionally, the hydrogen storage potential in India is uncertain due to the uncertainty in geological data. For these regions, alternatives to hydrogen salt cavern storage solutions are necessary to support the deployment of renewable energy systems. These could include other underground storage options, such as porous rock storage, which is expected to be more expensive and requires further research [21]. Another potential solution for these regions could be to explore alternative backup energy carriers, such as biomass or hydropower [47].

It is important to note that the end of life of hydrogen cavern storage is not fully understood. On the one hand, Berest et al. [13] estimate the lifetime of a salt cavern storage to be about 60 years, after which the cavern is typically abandoned due to salt creep. On the other hand, large-scale use of salt caverns started in the 1960s and caverns are still in operation. Therefore, it is necessary to consider the uncertainty of the lifetime to assess the sufficiency of hydrogen storage. In general, the sufficiency is greater than 1,000%, indicating that the potential of hydrogen storage lasts for a period exceeding 600 years. In terms of the energy system transformation until 2050, this does not pose any limitation on salt cavern usage. For continuous use of hydrogen salt caverns, new methods of hydrogen storage must be found or the potential for reusing existing caverns must be investigated in the long term.

This analysis addresses potential sources of uncertainty by establishing upper and lower bounds for geological suitability and an upper bound for the storage requirements for renewable energy systems. Accordingly, this methodology provides a reliable lower bound estimate of the sufficiency of hydrogen salt caverns for future renewable energy systems. Nevertheless, the estimation of global potentials of hydrogen salt caverns comes with some inevitable shortcomings. On one hand, there is an uncertainty in geological salt analysis. This can be expressed in terms of the uncertainty regarding the existence of undiscovered salt deposits. Based on mapped existing borehole data [48], the greatest probability for undiscovered salt deposits lies in regions with limited geological exploration, such as South America, Africa and Central Asia. Additionally, not all discovered salt deposits are fully geologically explored, resulting in partial understanding of their suitability. While the guaranteed suitable cavern can meet 45% of the global long-term storage need, further exploration would improve the knowledge base regarding suitable locations for salt caverns.

Finally, this study shows that the technical potential of hydrogen storage can play a major role in stabilizing 43-88% of future renewable electricity systems. To improve the deployment of hydrogen storage in salt caverns, we propose the following actions. At the national level, research on the national site-specific feasibility of salt caverns should be promoted, as stated by Lankof et al. [27] for Poland or in the US [49]. Also, a better understanding of the operation of salt caverns in renewable energy systems through pilot studies such as HySecure in the UK [50] or H2CAST Etzel in Germany [51]. Furthermore, the geographically limited occurrence of hydrogen salt cavern storage needs to be considered in the planning of infrastructure projects such as the Hydrogen Backbone [19] to create "hydrogen hubs" for the stabilization of energy systems as described by Caglayan [52]. Sectoral integration along the entire hydrogen value chain of production, storage, transport and consumption, especially for re-electrification can also reduce risks. For international cooperation, international partnerships, participation in international forums such as the International Hydrogen Trade Forum, and bilateral agreements can help to meet future storage needs.

## Methodology

For deriving the global potentials, three different steps are necessary. Initially, the geological suitable salt deposits are identified. Afterwards, a land eligibility analysis is conducted. Finally, the individual caverns are placed, and capacities are calculated. Emphasize is given to open data to allow sharing the obtained results openly.

### 3.1 Geological assessment

To determine the global hydrogen salt cavern potential, the suitable geological salt formations need to be determined. Consequently, four criteria must be met: a specific depth, a minimal thickness, a minimal size of the salt deposit as well as a maximum share of insoluble minerals need to be known (see table) [22]. In addition to that, the exact georeferenced location must be known for the eligibility analysis.

Table 1: Eligibility criteria for salt caverns based on Caglayan et al. [22].

| Criteria | Range | Reason |
|---|---|---|
| Depth | $\in$ [500 m, 2000 m] | Stability |
| Thickness | >200 m | Stability and tightness |
| Insoluble Minerals | <25% | Volume decrease and cavern shape |
| Minimal Size | >15km$^2$ | Minimal distance to edge of deposit |

The criteria are applied to the global compilation of salt deposits from the Solution Mining Research Institute [18], which contains 174 georeferenced deposit groups and is the most complete dataset for salt deposits with global coverage. To enhance the accuracy of the dataset, it was further updated with national data from the Netherlands [25, 53], the United Kingdom [24], the USA [49], Canada [54], and Poland [26], and checked against 42 studies listed in the Supplementary Data.

While local studies provide higher accuracy for specific regions, globally consistent data ensure methodological consistency and comparability across different areas. Because salt deposits can be geologically complex, a two-tier classification system is introduced: Deposits that meet the geological criteria are categorized as "guaranteed suitable", while "partially suitable" salt deposits are handled separately. Where possible, the 34 partially suitable deposits were further subdivided into suitable and unsuitable parts based on the geological data.

This approach recognizes the range of geological suitability, with the guaranteed suitable scenario representing the lower boundary and the combined guaranteed and partly suitable deposits marking an upper boundary. The actual suitability of all globally known deposits lies somewhere between these two end members. The detailed results of this suitability assessment can be found in the Supplementary Data.

### 3.2 Land eligibility analysis

Following the geological assessment, a land eligibility assessment is conducted to identify all suitable land areas for cavern placement. The approach from Caglayan et al. [22] is expanded globally by using the datasets from Franzmann et al. [46] with global coverage. Several criteria are taken into account. Firstly, due to ground subsidence, all settlements and buildings at a minimal distance of 2000 to 2500 m are excluded [22]. In addition, all other infrastructure, such as streets and power lines are not available for cavern placement based on Caglayan et al. [22]. Furthermore, in order to prevent damages to the caverns, active seismic faults are excluded at a distance of 200 m [22]. Based on Carneiro et al. [30], large airports are excluded at a distance of 20 km. Finally, in consideration of nature protection, no salt caverns can be built within protected areas [22]. Furthermore, as salt cavern construction utilizes water [55], regions with high water stress are not used for cavern placement. The complete list of criteria and their data sources can be found in the appendix.

The land eligibility analysis is conducted with the Python package GLAES [56] for land eligibility analysis at a resolution of 100mx100m for the 43 countries with suitable geological formations. Following this step, all suitable global areas for cavern placement are determined.

Furthermore, a further scenario is defined, which allows for directional drilling. In this scenario, the cavern can be placed up to 5000 m [57] below all areas apart from buildings and seismic faults, as the ground subsidence can also occur with horizontal drilling.

### 3.3 Placement and capacity estimation

The global suitable areas for hydrogen salt cavern placement are used to calculate the storage capacity. In the first step, location specific caverns are placed within suitable areas and with sufficient distance to each other. Based on thoughts from Caglayan et al. [22], two different salt cavern shapes are used for domal, and structural salt based on existing caverns (see Table 2). For placing the specific placements, a separation distance of 4 diameter is used for stability reasons [22]. Two additional technology scenarios are defined, where the placements are set a separation of 3 or 5 diameters.

**Table 2: Definition of cavern size based on Caglayan et al. [22]**

|                     | Bedded salt  | Domal salt  |
|---------------------|--------------|-------------|
| Height              | 120 m        | 300 m       |
| Diameter            | 84 m         | 58 m        |
| Volume              | 500,000 m³   | 750,000 m³  |
| Separation distance | 336 m        | 232 m       |

The maximal capacity of a salt cavern $E_{cap}$ is calculated based on the real gas equation of hydrogen and the minimal and maximum lithological pressure of the salt cavern $p_{min} = 0.3\ p_{max}$ and $p_{max} = 0.8\ p_{lith}$ [22]:

$$E_{cap} = H_U V_{Cavern}(1-\phi)\frac{\frac{p_{max}}{Z\ p_{max}} - \frac{p_{min}}{Z\ p_{min}}}{R/M_{H_2}T}$$

Here, $H_U$ is the lower heating value of hydrogen, $V_{Cavern}$ the cavern volume, $\phi$ the share on insoluble minerals, $R$ the ideal gas constant, $Z$ the compressibility factor, $M$ the molar mass of hydrogen and $T$ the cavern temperature. The proportion of insoluble minerals $\phi$ is assessed based on the deposit specific sources from the geological analysis (see Section 3.1). The lithological pressure is derived by $p_{lith} = \rho_{rock} g d_{LLC}$ [22] with $\rho_{rock} = 2550\ kg/m^3$ [58] and $d_{LCC}$ being the depth of the cavern top. In the end, a global shape file with all georeferenced placements with the respective hydrogen storage capacity is obtained.

## Acknowledgements

This work was supported by the Helmholtz Association under the program "Energy System Design".

This work was funded by the European Union (ERC, MATERIALIZE, 101076649). Views and opinions expressed are however those of the authors only and do not necessarily reflect those of European Union or the European Research Council Executive Agency. Neither the European Union nor the granting authority can be held responsible for them

## Authors contributions

**David Franzmann:** Conceptualization, Methodology, Software, Formal analysis, Investigation, Data Curation, Writing - Original Draft, Writing - Review & Editing, Visualization, Supervision, Project administration. **Thora Schubert:** Conceptualization, Methodology, Formal analysis, Review & Editing. **Heidi Heinrichs:** Conceptualization, Writing - Review & Editing, Resources, Supervision. **Peter Kukla:** Writing - Review & Editing, Supervision, **Detlef Stolten:** Writing - Review & Editing, Supervision.

## Supplementary data

The results for detailed placements as well as aggregated regions are attached to this publication.

## Appendix

**Table 3: Total Hydrogen Storage Potentials in Salt Caverns per Global Region**

|  | Capacity [TWh] |  | Hydrogen Storage Sufficiency [%] |  |
| --- | --- | --- | --- | --- |
| Region Name | Partially and guaranteed suitable | Guaranteed suitable | Partially and guaranteed suitable | Guaranteed suitable |
| North America | 354360 | 21216 | 36172 | 2166 |
| Mexico | 48996 | 1564 | 39364 | 1256 |
| Caribbean | 0 | 0 | 0 | 0 |
| SouthAmerica North | 31 | 31 | 30 | 30 |
| South America South | 0 | 0 | 0 | 0 |
| Brazil | 62240 | 0 | 37402 | 0 |
| South America North-East | 0 | 0 | 0 | 0 |
| Europe | 15380 | 10412 | 2044 | 1384 |
| Iceland | 0 | 0 | 0 | 0 |
| Pakistan + Afghanistan | 90 | 0 | 133 | 0 |
| Middle East | 18185 | 0 | 6895 | 0 |
| North Africa | 11952 | 7352 | 9987 | 6143 |
| West Africa | 0 | 0 | 0 | 0 |
| Central Africa | 19318 | 0 | 138785 | 0 |
| East Africa | 523 | 523 | 524 | 524 |
| South Africa | 29172 | 2118 | 23983 | 1741 |
| Russia + surroundings | 47643 | 87 | 15987 | 29 |
| China + North Korea | 43074 | 43074 | 4086 | 4086 |
| Japan + South Korea | 0 | 0 | 0 | 0 |

| India + surroundings | 9995 | 0 | 1529 | 0 |
| Iran + Afghanistan | 6671 | 6671 | 7446 | 7446 |
| South-East Asia | 0 | 0 | 0 | 0 |
| Australia | 119677 | 119677 | 167777 | 167777 |
| Micronesia | 0 | 0 | 0 | 0 |
| Melanesia | 0 | 0 | 0 | 0 |
| Polynesia | 0 | 0 | 0 | 0 |
| Central American Common Market | 1156 | 0 | 3321 | 0 |
| Antarctica | 0 | 0 | 0 | 0 |